\documentclass{article}

\PassOptionsToPackage{numbers, compress}{natbib}
\usepackage[preprint]{neurips_2026}

\usepackage[utf8]{inputenc} % allow utf-8 input
\usepackage[T1]{fontenc}    % use 8-bit T1 fonts
\usepackage{hyperref}       % hyperlinks
\usepackage{url}            % simple URL typesetting
\usepackage{booktabs}       % professional-quality tables
\usepackage{amsfonts}       % blackboard math symbols
\usepackage{nicefrac}       % compact symbols for 1/2, etc.
\usepackage{microtype}      % microtypography
\usepackage{graphicx}
\usepackage{xcolor}         % colors
\usepackage{amsmath}

\title{Triple Configuration of Brain Networks Based on Recurrent Neural Networks: The Synergistic Effects of Exogenous Stimuli, Task Demands, and Spontaneous Activity}

\author{%
  Binghao Yang \\
  School of Future Technology, University of Chinese Academy of Sciences \\
  Beijing 101408, China \\
  \texttt{yangbinghao@xwhosp.org} \\
  \And
  Guangzong Chen \\
  Department of Electrical and Computer Engineering \\
  University of Pittsburgh \\
  Pittsburgh, PA 15260, USA \\
  \texttt{guangzong@pitt.edu} \\
}
\begin{document}

\maketitle

\begin{abstract}
	% In the natural condition, the functional structure and activity of the brain network can be dynamically configured by both the external environments and internal states, which is the foundation of the cognitive flexibility and higher-order intelligence. By means of the RNN model with neural dynamic constraint from individual source-localized resting-state EEG of 114 participants, this study aims to provide a comprehensive exploration to clarify triple brain network configurations from exogenous and endogenous factors, involving external stimuli, information processing tasks and spontaneous activities. Detailed comparisons determine the key role of the parietal network in supporting multiple types of configuration patterns. Further, the anterior/posterior parietal regions exhibit functional specialization under such multiple configurations from different modalities of stimulus input. This work tries to propose a triple configuration framework, separates different potential factors and underscores the significance of parietal regions under the brain network configuration. 

	The foundation of cognitive flexibility and higher-order intelligence lies in the functional structure and
	activity of brain networks, which can be dynamically configured by both external environments and internal
	states.
	However, decoding these dynamics from high-dimensional neural data remains a challenge.
	In this study, we propose a computational framework using Recurrent Neural Networks (RNNs) with neural dynamic
	constraints to model source-localized resting-state EEG data from $114$ participants.
	We aim to clarify the ``triple brain network configurations'' driven by exogenous and endogenous factors, including external stimuli, information processing tasks, and spontaneous activities.
	Our model identifies the parietal network as a critical hub supporting these multiple configuration patterns.
	Furthermore, we reveal that the anterior and posterior parietal regions exhibit distinct functional specializations under different stimulus modalities.
	By formalizing a triple configuration framework, this work separates latent factors of brain dynamics and underscores the computational significance of parietal regions in orchestrating higher-order intelligence.
\end{abstract}

\section{Introduction}
% \color{red}

% As a dynamic system, the brain receives and processes successive external information, as well as performs multiple cognitive functions dynamically~\citep{avena2018communication, petersen2015brain}.
% The brain exhibits highly complex activity patterns for the response, encoding, and processing of
% information~\citep{chialvo2010emergent}, and can change patterns between distinct cognitive tasks and contexts~\citep{medaglia2018functional, mattar2016flexible, lam2025prefrontal}.
% The functions of such cognitive processing with external information inputs rely on the recruitment of multiple brain regions, i.e., the configuration of brain networks~\citep{shine2018principles}.
% Determining which brain regions participate in information processing based on cognitive requirements is the foundation of the complexity and flexibility of the brain.

As a dynamic system, the brain continuously receives and processes external information while performing multiple cognitive functions~\citep{avena2018communication, petersen2015brain}. It exhibits highly complex activity patterns for encoding and processing information~\citep{chialvo2010emergent}, and can flexibly transition between distinct cognitive tasks and contexts~\citep{medaglia2018functional, mattar2016flexible, lam2025prefrontal}. Such cognitive processing relies on the recruitment of distributed brain regions, effectively manifesting as the dynamic configuration of brain networks~\citep{shine2018principles}. The ability to selectively recruit brain regions to meet varying cognitive demands serves as the foundation of the brain's complexity and flexibility.

Specifically, the configuration of brain networks is crucial for the processing of complex information and higher-order cognitive functions across multiple domains, including sensory processing, attention, learning, and memory. For instance, interactions between the insula and temporal networks enable the brain to decompose information within complex auditory environments~\citep{altmann2008effects, bidet2007effects}. Similarly, attention modulates visual network configuration to amplify the encoding of specific features in visual input~\citep{henderson2025dynamic}, while learning processes in uncertain environments involve the reconfiguration of the frontoparietal network~\citep{li2023functional, li2024amygdala}. Furthermore, the configuration within the limbic network—specifically the interaction between the hippocampus and the amygdala—supports memory processing distinctively during the encoding and retrieval phases~\citep{bassett2011dynamic, kao2020functional}.

However, several key questions remain elusive. Given the complexity of cognition, network configuration is influenced by multifarious factors. For example, the concurrent presence of primary sensory responses and higher-order cognitive processing following task-related stimuli makes it difficult to disentangle their respective contributions to network configuration. Isolating these factors is essential for clarifying the specific roles of brain networks in stimulus response versus cognitive processing. Additionally, in natural environments, brain networks must accommodate multiple inputs and tasks, a process further modulated by the brain's intrinsic state~\citep{shine2018principles}. The network mechanisms enabling adaptation to such diverse demands, as well as their relationship with spontaneous brain activity, remain unclear.

% 	In this study, using recurrent neural networks (RNN) to model neural dynamics from resting-state electroencephalography (EEG) and imposing different types of external inputs and information processing tasks on such neural-dynamic-constraint models, we proposed a triple configuration framework from exogenous and endogenous effects on the brain network.
% The first type of configuration originated from the intuitive brain network response to external inputs (auditory/visual cortex inputs) (Figure~\ref{fig:overview}A, Left);
% The second type of configuration originated from the internal information organization based on the task requirements (Figure~\ref{fig:overview}A, Middle);
% And the third type of configuration originated from the spontaneous brain activities, independent from external stimuli and task but associated with the brain state (Figure~\ref{fig:overview}A, Right).
% The parietal network simultaneously participated in the configurations by external stimuli, tasks and spontaneous activities.
% 	By comparing the configuration patterns under different modalities of sensory inputs, this study further demonstrated the emerged specialization of anterior/posterior parietal network in the network configurations corresponding to the auditory/visual input.
% 	These results revealed the multiple effects of endogenous and exogenous factors on brain network configurations, revealing the hub role of the parietal network in them.
% \textcolor{blue}{
In this study, by employing recurrent neural networks (RNNs) to model neural dynamics from resting-state EEG and subjecting these biologically constrained models to various external inputs and cognitive tasks, we propose a ``triple configuration framework'' governing brain networks via exogenous and endogenous effects. For a detailed description of the study pipeline and the triple configuration framework, refer to Appendix A (Figure~\ref{fig:overview}).
% }

\color{black}
\section{Materials and Methods}

\subsection{Participants, EEG Recording, Preprocessing, and Source Localization}

We recruited $114$ participants for this study. EEG data were recorded using a $64$-channel system and preprocessed using standard pipelines, including filtering and Independent Component Analysis (ICA) for artifact removal. Detailed information regarding participants, data acquisition, preprocessing, and source localization is provided in Appendix A.
\subsection{Recurrent Neural Network Model Training}

\begin{figure}
	\centering
	\includegraphics[width=\linewidth]{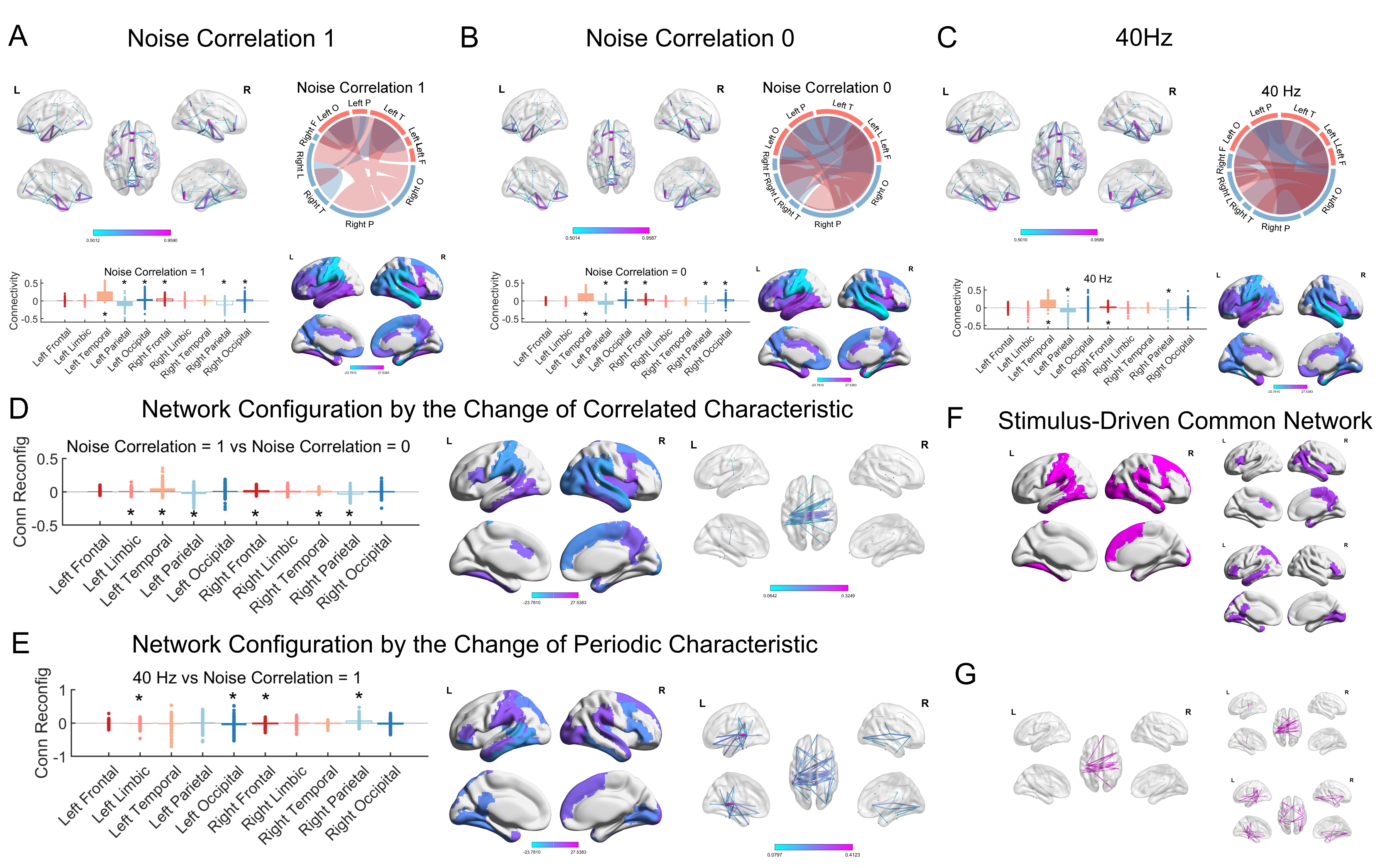}
	\caption{Stimulus-driven network configuration under auditory cortex input.
		(A) Functional connectivity patterns under white noise input with bilateral correlation
		1. Upper-Left: pairwise whole brain connectivity; Upper-right: sub-network connectivity; Lower-left: averaged seed-point connectivity of sub-networks. Lower-right: seed-point connectivity mapping on the whole brain.
		(B) Same as (A) under white noise input with bilateral correlation 0.
		(C) Same as (A) under $40$~Hz input.
		(D) Network configuration by the change of correlated characteristic (Noise correlation 1 vs Noise correlation 0). Left: seed-point connectivity configuration of sub-networks; Middle: seed-point connectivity configuration mapping on the whole brain; Right: pairwise connectivity configuration of the whole brain.
		(E) Same as (D) by the change of periodic characteristic (Noise correlation 1 vs $40$~Hz).
		(F) Left: Stimulus-driven common network by combining (D) and (E); Upper-right: Network only configured by the change of correlated characteristic; Lower-right: Network only configured by the change of periodic characteristic. (G) Same as (F) shown at the pairwise connectivity level. * $p <0.05$ after Bonferroni correction.}
	\label{fig:stimulus_driven}
\end{figure}

\begin{figure}
	\centering
	\includegraphics[width=\linewidth]{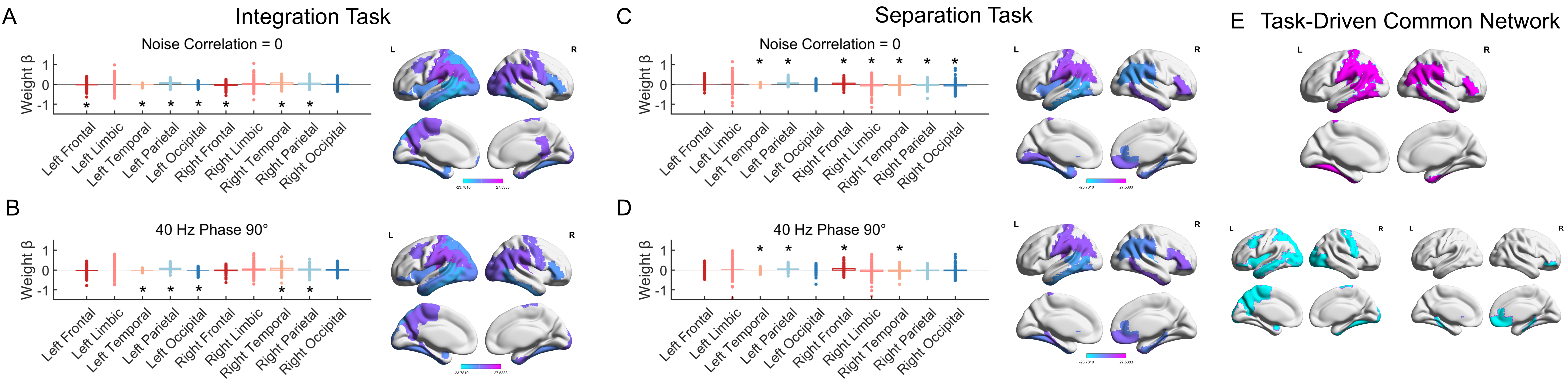}
	\caption{Task-driven network configuration under auditory cortex input. (A) Network configuration performing integration task under the input of white noise with correlation 0. Left: seed-point connectivity configuration of sub-networks; Right: seed-point connectivity configuration mapping on the whole brain. (B) Same as (A) under the input of $40$~Hz with phase lag $90^\circ$. (C) Network configuration performing separation task under the input of white noise with correlation 0. Left: seed-point connectivity configuration of sub-networks; Right: seed-point connectivity configuration mapping on the whole brain. (D) Same as (C) under the input of $40$~Hz with phase lag $90^\circ$. (E) Upper: Task-driven common network by combining (A) - (D); Lower-left: Network only configured by the integration task; Lower-right: Network only configured by the separation task. * $p <0.05$ after Bonferroni correction.
	}
	\label{fig:task_driven}
\end{figure}

\begin{figure}
	\centering
	\includegraphics[width=\linewidth]{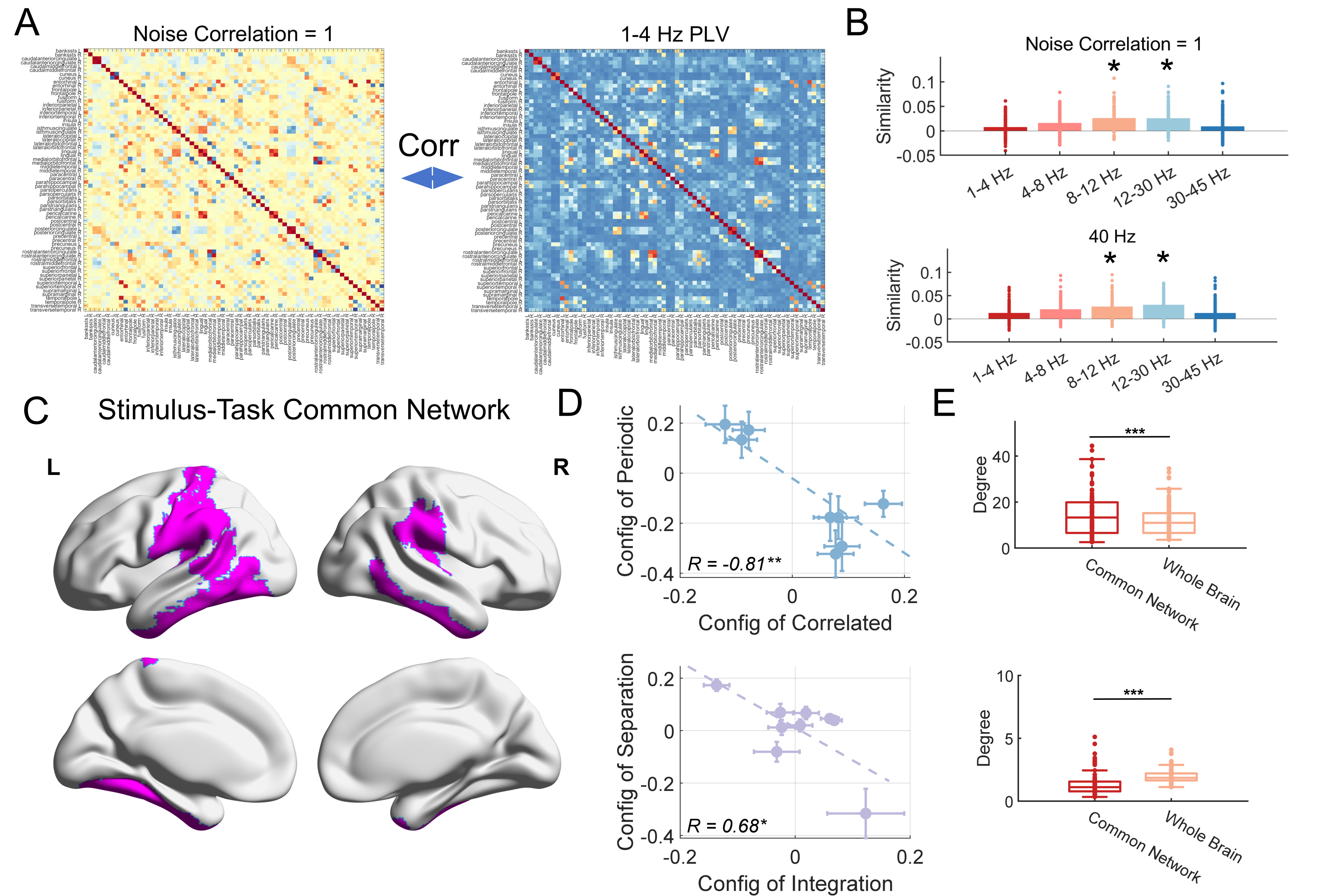}
	\caption{Spontaneous-activity network configuration under auditory cortex input.
		(A) Schematic of similarity analysis, i.e. calculating the correlation coefficient between stimulus-driven functional connectivity and resting-state functional connectivity.
		(B) Similarity analysis across five frequency bands: delta ($1-4$~Hz), theta ($4-8$~Hz), alpha ($8-12$~Hz), beta ($12-30$~Hz), and gamma ($30-45$~Hz). Upper: Similarity with functional connectivity under white noise input. Lower: Similarity with functional connectivity under $40$~Hz input.
		(C) Stimulus-task common network combining Figure~\ref{fig:stimulus_driven}(F) and Figure~\ref{fig:task_driven}(E).
		(D) Correlation of common network configuration patterns. Upper: different types of stimulus input; Lower: different types of tasks.
		(E) Comparison between the degree of brain regions in the common network and the whole-brain averaged degree. Upper: alpha band; Lower: beta band. * $p <0.05$ after Bonferroni correction; *** $p <0.001$.
	}
	\label{fig:spontaneous}
\end{figure}
\begin{figure}
	\centering
	\includegraphics[width=\linewidth]{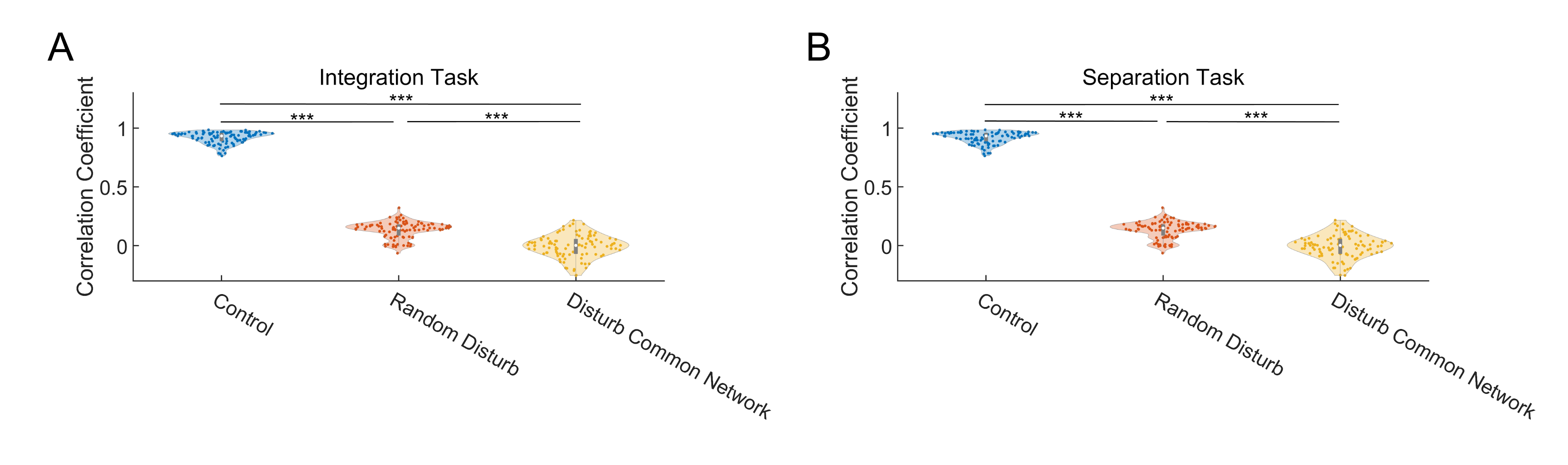}
	\caption{Elimination experiment. (A) Change of the model performance (measured by the correlation coefficient between model output signal and target signal) under control condition (without elimination), random-disturbed condition (eliminating read-out weights randomly) and common network disturbed condition (eliminating read-out weights corresponding with the common network) in the integration task. (B) Same as (A) in the separation task. *** $p <0.001$ after Bonferroni correction; }
	\label{fig:elimination}
\end{figure}

\begin{figure}
	\centering
	\includegraphics[width=\linewidth]{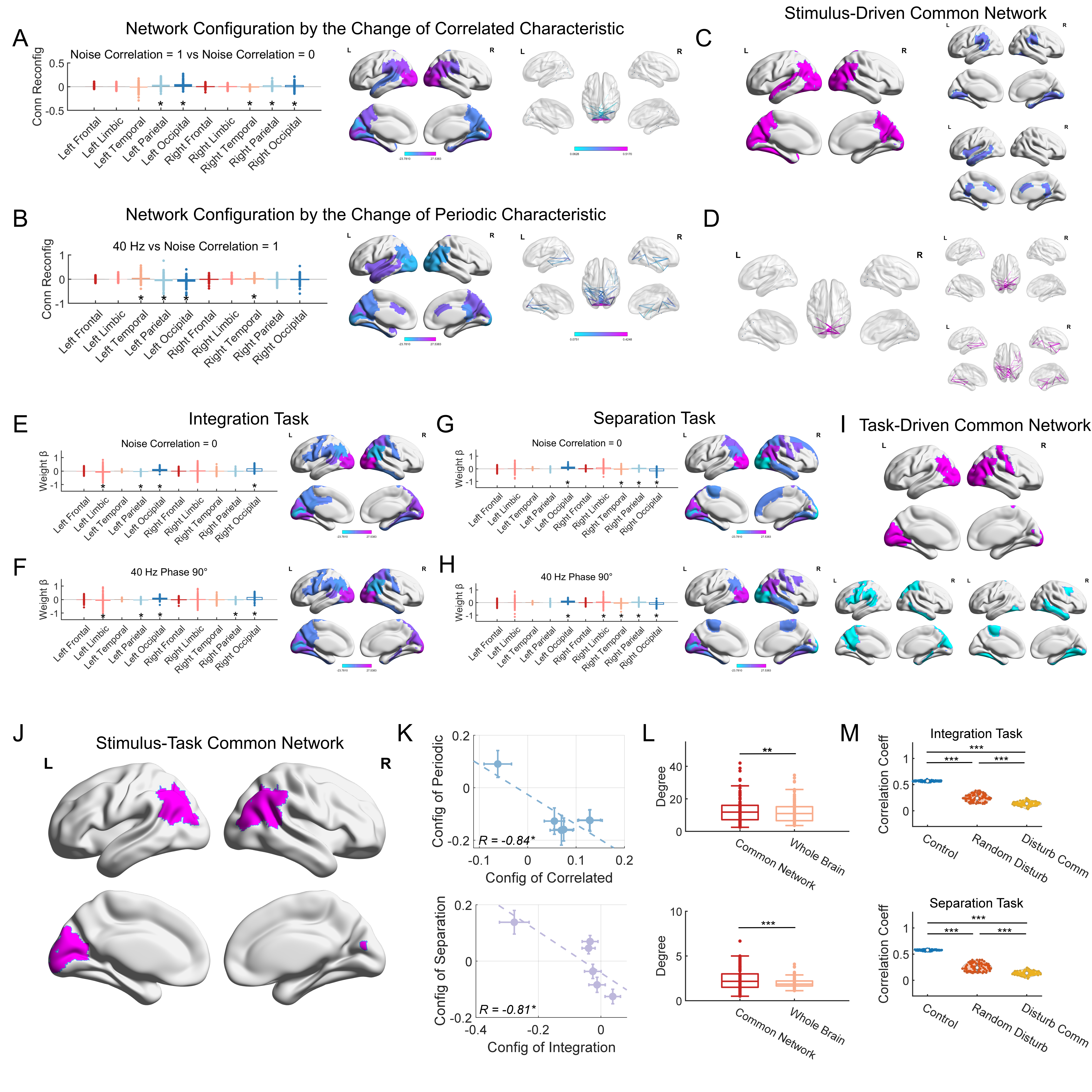}
	\caption{Triple network configuration under visual cortex input. (A)-(D) Stimulus-driven configuration. (A) Network configuration by the change of correlated characteristic (Noise correlation 1 vs Noise correlation 0). Left: seed-point connectivity configuration of sub-networks; Middle: seed-point connectivity configuration mapping on the whole brain; Right: pairwise connectivity configuration of the whole brain. (B) Same as (A) by the change of periodic characteristic (Noise correlation 1 vs 40 Hz). (C) Left: Stimulus-driven common network by combining (A) and (B); Upper-right: Network only configured by the change of correlated characteristic; Lower-right: Network only configured by the change of periodic characteristic. (D) Same as (C) shown at the pairwise connectivity level. (E)-(I) Task-driven configuration. (E) Network configuration performing integration task under the input of white noise with correlation 0. Left: seed-point connectivity configuration of sub-networks; Right: seed-point connectivity configuration mapping on the whole brain. (F) Same as (E) under the input of 40Hz with phase lag 90°. (G) Network configuration performing separation task under the input of white noise with correlation 0. Left: seed-point connectivity configuration of sub-networks; Right: seed-point connectivity configuration mapping on the whole brain. (H) Same as (G) under the input of 40Hz with phase lag 90°. (I) Upper: Task-driven common network by combining (E) - (H); Lower-left: Network only configured by the integration task; Lower-right: Network only configured by the separation task. (J)-(M) Spontaneous-activity configuration. (J) Stimulus-task common network combining (C) and (I). (K) Correlation of common network configuration patterns. Upper: different types of stimulus input; Lower: different types of tasks. (L) Comparison between the degree of brain regions in the common network and the whole-brain averaged degree. Upper: alpha band; Lower: beta band. (M) Elimination experiment. Upper: integration task; Lower: separation task. * p <0.05, **p<0.01, ***p<0.001 after Bonferroni correction.}
	\label{fig:visual_triple}
\end{figure}

We trained a Recurrent Neural Network (RNN) to predict individual source-localized EEG signals at time $t+1$ given the input at time $t$. The RNN model comprised a fully connected population of $200$ hidden units receiving a $68 \times 1$ input vector, where each dimension corresponded to the signal from a single brain region. The hidden unit activity was governed by:
\begin{equation}
	h_t = f(W_{ih} x_t + b_{ih} + W_{hh} h_{t-1})
\end{equation}
where $f(\cdot)$ is the hyperbolic tangent activation function ($\tanh$), $h_t \in \mathbb{R}^{200}$ represents the hidden state, $x_t \in \mathbb{R}^{68}$ is the input EEG signal at time $t$, $W_{hh} \in \mathbb{R}^{200 \times 200}$ is the recurrent weight matrix, and $W_{ih} \in \mathbb{R}^{200 \times 68}$ is the input weight matrix. The output was computed as:
\begin{equation}
	y_{\text{pred}}(t) = W_{\text{out}} h_t
\end{equation}
where $y_{\text{pred}}(t) \in \mathbb{R}^{68}$ denotes the predicted EEG signal at time $t+1$, and $W_{\text{out}} \in \mathbb{R}^{68 \times 200}$ is the output weight matrix.
EEG data were divided into $10$-second segments ($2500$ data points) with $75\%$ overlap; segments with significant artifacts were excluded. At time $t$, given the EEG signals in the interval $[t, t+2500]$, the RNN model was trained to predict the EEG signals at the next time step. This training procedure allowed the RNN to learn the intrinsic neural dynamics of the EEG signals, providing a biologically plausible model for analyzing triple brain network configurations.

Prior to training, all model parameters were initialized with random values drawn from a standard Gaussian distribution $\mathcal{N}(0,1)$. The training batch size was set to $64$, and the model was trained for $4000$ iterations using Backpropagation Through Time (BPTT)~\citep{lillicrap2019backpropagation} with the Adam optimizer (learning rate $2 \times 10^{-3}$). The objective was to minimize the Mean Squared Error (MSE) across all training data segments:
\begin{equation}
	\text{MSE} = \frac{1}{N} \sum_{i=1}^{N} \sum_{t=1}^{T} (y_i(t) - y_{\text{pred},i}(t))^2
\end{equation}
where $N$ indicates the batch size ($64$). $y_i(t)$ represents the ground truth EEG signal, and $y_{\text{pred},i}(t)$ denotes the predicted signal. The model implementation and parameter updates were performed using PyTorch~\citep{paszke2019pytorch}.

\subsection{Stimulus Design}
To simulate different sensory modalities, we designated the bilateral transverse temporal regions (auditory cortex) as auditory input nodes and the bilateral lateral occipital regions (visual cortex) as visual input nodes. We examined the impact of stimulus characteristics by generating four stimulus types: (1) aperiodic white noise with perfect bilateral correlation ($r=1$); (2) uncorrelated aperiodic white noise ($r=0$); (3) periodic $40$~Hz signals with zero phase lag; and (4) periodic $40$~Hz signals with a $90^\circ$ phase lag. Our analysis focused on two dimensions: periodic characteristics (white noise vs. $40$~Hz) and correlation characteristics (correlated vs. uncorrelated).

\subsection{Stimulus-Driven Network Configuration Analysis}
% TODO: @ybh: check this part
% First we analyzed the impact of stimulus characteristics on the brain network configuration (Figure 1(E)). For auditory inputs, stimuli were imposed on the RNN input channels corresponding with the bilateral transversetemporal regions; for visual inputs, stimuli were imposed on the input channels corresponding with the bilateral lateraloccipital regions; for multi-modal auditory-visual inputs, stimuli were imposed on both the transversetemporal regions and lateraloccipital regions. The output signals of the RNN model reflected the response to the stimuli under the neural dynamic constraints modeled from EEG signals. The functional connectivity under specific stimulus was evaluated as the pairwise Pearson correlation coefficients of the RNN output signals. At the level of brain regions, we took the regions of stimulus input as seed points and calculated the functional connectivity between seed points and other brain regions. The seed-point connectivity patterns under white noise with correlation 1 and correlation 0 / white noise and 40Hz were compared to identify the brain regions with significant change of connectivity, determining the brain networks configured by the correlated/periodic characteristic. The stimulus-driven common network comprised brain regions configured by the change of both the correlated characteristic and periodic characteristic.
We analyzed the impact of stimulus characteristics on brain network configuration. Stimuli were applied to RNN input channels corresponding to auditory, visual, or combined regions. The RNN output reflected the network response under learned neural dynamic constraints. Functional connectivity was evaluated using pairwise Pearson correlation coefficients of the output signals. We computed seed-based connectivity from the input regions to the rest of the brain. By comparing connectivity patterns between different stimulus conditions (e.g., correlated vs. uncorrelated noise), we identified brain regions exhibiting significant connectivity changes. The ``stimulus-driven common network'' was defined as the set of regions showing configuration changes under both periodic and correlation manipulations.

\subsection{Task-Driven Network Configuration Analysis}
We further investigated network configuration under specific information processing tasks. A readout layer ($68 \times 1$) was added to the RNN to generate a scalar output $y_{\text{target}}$:
\begin{equation}
	\text{Output} = \beta^T y_{\text{pred}}
\end{equation}
where $\beta$ represents the readout weights, interpretable as regional activation. Two tasks were defined: (1) Integration, where the target was the sum of bilateral inputs; and (2) Separation, where the target was the difference. Readout weights were optimized using least squares. Significant weights identified task-specific activation patterns. The ``task-driven common network'' comprised regions activated by both integration and separation tasks. Combining this with the stimulus-driven network yielded the ``stimulus-task common network''.

\subsection{Spontaneous-Activity Network Configuration Analysis}
We computed the Phase Locking Value (PLV)~\citep{aydore2013note, sazonov2009investigation} of resting-state EEG to characterize spontaneous functional connectivity. Signals were band-pass filtered into five frequency bands (delta to gamma). The PLV between signals $s_1(t)$ and $s_2(t)$ was calculated using their analytic signals $z_i(t)$ obtained via the Hilbert transform:
\begin{equation}
	\text{HT}(s_i(t)) = \frac{1}{\pi} \int_{-\infty}^{\infty} \frac{s_i(\tau)}{t - \tau} d\tau
\end{equation}
The phase difference was computed as:
\begin{equation}
	\Delta\phi(t) = \arg\left(\frac{z_1(t) z_2^*(t)}{|z_1(t)| |z_2(t)|}\right)
\end{equation}
where * denotes the complex conjugate and $|\cdot|$ represents the modulus of the complex number. The PLV was defined as:
\begin{equation}
	\text{PLV} = \left|\frac{1}{N} \sum_{t=1}^{N} e^{j \Delta\phi(t)}\right|
	\label{eq:PLV}
\end{equation}
% \color{red}
% 这上面我看了改了
PLV ranges from 0 to 1, where 0 represents no phase synchrony and 1 represents perfect phase synchrony. The PLV was calculated in five frequency bands: delta ($1-4$~Hz), theta ($4-8$~Hz), alpha ($8-12$~Hz), beta ($12-30$~Hz), and gamma ($30-45$~Hz).

% We assessed the similarity between stimulus-driven connectivity and spontaneous PLV matrices using correlation analysis. A graph was constructed from the PLV matrix (threshold $>0.6$), and the degree of nodes within the stimulus-task common network was computed to evaluate the influence of spontaneous activity (Figure~\ref{fig:overview}G).

We compared the similarity between the stimulus-configured network and the spontaneous activity network by calculating the correlation coefficient between the functional connectivity matrix under the white noise/$40$~Hz input and the spontaneous PLV matrix. Then we constructed the connectivity graph from the PLV matrix with significant similarity. The nodes of the graph correspond to the brain regions, and the connections greater than $0.6$ in the PLV matrix were conserved as the edges of the graph. We calculated the degree of nodes corresponding to the brain regions in the stimulus-task common network to illustrate the role of the spontaneous activity configuration on the common network.

\subsection{Analysis Under Different Brain States}
As a supplementary analysis, we compared the stimulus-task configuration patterns between the drowsiness state (postprandial somnolence/sleep deprivation) and the wakefulness state to assess the influence of brain state changes on brain network configuration. To do this, we trained an individual RNN model using EEG data recorded during postprandial somnolence or sleep deprivation. All other analysis procedures remained the same to obtain the stimulus and task configuration network patterns during the drowsiness state.

\section{Statistical Analysis}

For the analysis of stimulus-driven network configurations, we compared the group-level seed-point connectivity against $0$ at both the sub-network and single-region levels to determine which brain regions responded to the stimuli. To identify regions modulated by changes in stimulus characteristics, the group-level difference in seed-point connectivity between pairs of stimuli was also tested against $0$. For the task-driven analysis, the group-level $\beta$ was compared against $0$ to map the brain regions engaged by the tasks. All comparisons were conducted using $t$-tests, with statistical significance set at $p < 0.05$ following a Bonferroni correction. Finally, to compare functional connectivity within and outside the common network, as well as task performance with and without elimination of the common network, we conducted a one-way ANOVA followed by post-hoc two-sided paired $t$-tests (significance set at $p < 0.05$, Bonferroni-corrected).

\section{Experimental Results}

\subsection{Stimuli on the Auditory Cortex With Different Characteristics Configure Brain Connectivity Patterns}

Based on the RNN model with EEG dynamic constraints, we identified network configuration patterns under varying stimuli and tasks. We first examined stimulus-driven configuration under auditory input (bilateral transversetemporal regions). Three bilateral input types were analyzed: perfectly-correlated white noise ($r=1$), uncorrelated white noise ($r=0$), and periodic $40$~Hz input (with $0^\circ$ or $90^\circ$ phase lag). Functional connectivity across brain regions revealed similar significant patterns across stimulus types (Figure~\ref{fig:stimulus_driven}(A)-(C), Upper-Left, t-test, $p < 0.05$, Bonferroni correction). Seed-point connectivity mapping from the auditory cortex demonstrated extensive connections, primarily involving temporal and parietal regions (Figure~\ref{fig:stimulus_driven}(A)-(C), Bottom, t-test, $p < 0.05$, Bonferroni correction).

We next analyzed connectivity changes relative to stimulus characteristics: correlation ($r=1$ vs. $r=0$) and periodicity ($r=1$ noise vs. $40$~Hz). Correlation changes configured connectivity in the left limbic, right frontal, bilateral temporal, and parietal networks (Figure~\ref{fig:stimulus_driven}(D) left, middle, t-test, $p < 0.05$, Bonferroni correction). Periodicity changes configured the left limbic, right occipital, right frontal, and right parietal networks (Figure~\ref{fig:stimulus_driven}(E) left, middle, t-test, $p < 0.05$, Bonferroni correction). Modifying periodic frequency ($40$~Hz vs. $1$~Hz, Figure S1(A)) or phase lag ($0^\circ$ vs. $90^\circ$, Figure S1(B)) did not significantly alter connectivity (t-test, Bonferroni correction). Pairwise analysis showed correlation changes primarily modulated cross-hemispheric connectivity originating from temporal regions (Figure~\ref{fig:stimulus_driven}(D) Right, paired t-test, $p < 0.05$, Bonferroni correction), whereas periodicity changes modulated both cross- and intra-hemispheric connectivity (Figure~\ref{fig:stimulus_driven}(E) Right, paired t-test, $p < 0.05$, Bonferroni correction). A ``stimulus-driven common network'', responsive to both characteristic changes, was identified across frontal, temporal, and parietal regions (Figure~\ref{fig:stimulus_driven}(F) Left), modulating cross-hemispheric connections (Figure~\ref{fig:stimulus_driven}(G) Left). Thus, stimulus properties flexibly configure network connectivity across multiple regions.

\subsection{Different Auditory Information Processing Tasks Configure Brain Activation Patterns}

Next, we investigated task-driven network configurations by adding a $68 \times 1$ readout layer to the RNN, generating a 1D signal for task requirements. Readout weights $\beta$, derived via least squares, represented regional activation. We evaluated two tasks: integration (summing bilateral inputs) and separation (subtracting bilateral inputs). Inputs comprised uncorrelated white noise and $40$~Hz signals with $90^\circ$ phase lag. Activation mapping revealed that both tasks engaged distributed networks, primarily involving the left temporal and parietal, and right frontal, temporal, and parietal networks (Figure~\ref{fig:task_driven}(A)-(D), t-test, $p < 0.05$, Bonferroni correction). These patterns were task-specific and invariant to stimulus type (Figure S2, t-test, Bonferroni correction). By merging these patterns, we identified a ``task-driven common network'' localized to the temporal and parietal regions (Figure~\ref{fig:task_driven}(E), upper).

\subsection{The Stimulus-Task Common Network Under Auditory Input Is Coupled With the Hub of the Spontaneous Brain Connectivity Network, Forming a Triple-Configured Common Network}

We evaluated the coupling of this task-stimulus common network with spontaneous brain activity using Phase-Locking Value (PLV) (Figure~\ref{fig:spontaneous}(A)). Stimulus-driven connectivity under correlated white noise or $40$~Hz input highly correlated with alpha ($8-12$~Hz) and beta ($12-30$~Hz) PLV (Corrected $p < 1e-10$, Figure~\ref{fig:spontaneous}(B)). Combining stimulus and task common networks yielded a ``stimulus-task common network'' encompassing the anterior parietal and temporal regions (postcentral, supramarginal, bankssts, inferiortemporal, and fusiform) (Figure~\ref{fig:spontaneous}(C)). Configuration within this network was diametrically opposed across different conditions. Correlational characteristic changes inversely correlated with periodic characteristic changes (Pearson $R = -0.81, p < 0.01$, Figure~\ref{fig:spontaneous}(D), Upper). Similarly, integration activation inversely correlated with separation activation (Pearson $R = -0.68, p = 0.04$, Figure~\ref{fig:spontaneous}(D), Lower). In the alpha band, common network regions exhibited significantly higher degrees than the whole-brain average, indicating a hub role (Figure~\ref{fig:spontaneous}(E), Upper, paired t-test, $p < 0.05$); conversely, beta-band degree was lower (Figure~\ref{fig:spontaneous}(E), Lower, paired t-test, $p < 0.05$). Thus, this anterior parietal-temporal network undergoes triple configuration via external stimuli, tasks, and internal spontaneous activity.

Furthermore, state changes (drowsiness via postprandial somnolence or sleep deprivation) modulated network dynamics. RNNs trained on drowsiness EEG revealed altered connectivity and activation, predominantly in the supramarginal region, compared to wakefulness (Figure S3, S4, paired t-test, $p < 0.05$). This corroborates the influence of internal brain states on network configuration.

\subsection{Eliminating the Common Network Disrupted Task Performance}
To assess functional significance, we eliminated the common network's contribution by zeroing its corresponding readout weights during tasks. Performance was measured via output-target correlation. In both integration (Figure~\ref{fig:elimination}(A)) and separation (Figure~\ref{fig:elimination}(B)) tasks, eliminating the common network significantly reduced performance compared to normal and random-elimination conditions (One-way ANOVA, $p < 0.001$; post-hoc paired t-test, $p < 0.001$, Bonferroni correction). This underscores the network's critical role in information processing.
\subsection{The Triple-Configured Common Networks Under Auditory and Visual Stimuli Are Separated in the Parietal Region}

We then analyzed visual input at bilateral lateraloccipital regions. Seed-point connectivity changes due to correlation modulated the right temporal, and bilateral occipital and parietal networks (Figure~\ref{fig:visual_triple}(A) left, middle, t-test, $p<0.05$, Bonferroni correction). Periodicity changes modulated the left temporal, occipital, and parietal networks (Figure~\ref{fig:visual_triple}(B) left, middle, t-test, $p<0.05$, Bonferroni correction). Pairwise analysis showed both predominantly affected intra-parietal-occipital connectivity (Figure~\ref{fig:visual_triple}(A)-(B) Right, t-test, $p<0.05$, Bonferroni correction). A stimulus-driven common network was distributed across parietal and occipital regions (Figure~\ref{fig:visual_triple}(C) Left), modulating bilateral occipital connectivity (Figure~\ref{fig:visual_triple}(D) Left).

Task-driven patterns under visual input (uncorrelated noise and $40$~Hz with $90^\circ$ phase lag) activated temporal, occipital, and parietal networks (Figure~\ref{fig:visual_triple}(E)-(H), t-test, $p<0.05$, Bonferroni correction). The task-driven common network localized to occipital and parietal regions (Figure~\ref{fig:visual_triple}(I), upper).

Combining these yielded a visual stimulus-task common network (Figure~\ref{fig:visual_triple}(J)) in posterior parietal and occipital regions (inferiorparietal, cuneus, pericalcarine), anatomically distinct from the auditory common network. Dynamically, however, they were similar: stimulus correlation and periodicity changes were negatively correlated (Pearson $R = -0.84, p=0.04$, Figure~\ref{fig:visual_triple}(K), Upper), as were integration and separation activations (Pearson $R = -0.81, p=0.05$, Figure~\ref{fig:visual_triple}(K), Lower). The network exhibited significantly higher degree in alpha and beta bands, confirming spontaneous activity coupling (Figure~\ref{fig:visual_triple}(L), paired t-test, $p<0.05$). Eliminating the common network significantly impaired performance in both tasks compared to controls (Figure~\ref{fig:visual_triple}(M), One-way ANOVA, $p<0.001$; post-hoc paired t-test, $p<0.001$, Bonferroni correction). These findings demonstrate functional specialization within the parietal network: anterior regions process auditory information, while posterior regions process visual information.

\section{Discussion}
In this study, using an RNN model with source-localized EEG dynamic constraints, we separated and clarified triple brain network configurations modulated by external stimuli, tasks, and spontaneous brain activities. We identified the parietal regions as a common network serving different configuration types, distinguishing varying stimuli and task requirements through opposite configuration patterns. Furthermore, we demonstrated the functional specialization of the anterior and posterior parietal regions in network configurations under auditory and visual inputs, respectively.

Brain network configurations driven by stimuli or tasks occur across multiple levels---from regional activation \citep{jordan2001cortical, albouy2022supramodality, albouy2017selective} and local interactions to global networks supporting complex cognition \citep{mitsuhashi2022temporally, jimura2010age, li2019transitions}. These patterns are also influenced by internal states like sleep \citep{jang2024measuring, li2018comparison}. By integrating exogenous and endogenous effects, we pinpointed the parietal region as a crucial hub bridging these configurations.

Graph theoretical tools offer key insights into brain organization \citep{bullmore2009complex, wuthric2024neural, seguin2023brain}, highlighting optimal network properties for information processing, such as small-world \citep{watts1998collective, uehara2014efficiency, hallquist2018graph} and scale-free structures \citep{he2014scale, lynn2024emergent}. These theories underscore the importance of hub nodes for information transmission \citep{stam2024hub, van2013network}. Our findings confirm that parietal regions act as hub nodes within the spontaneous resting-state network, coupling with both stimulus- and task-driven configurations in complex environments.

Previous work indicates functional specialization within the parietal cortex: the anterior part primarily handles sensory information \citep{yao2023transformation}, whereas the posterior part manages higher-order functions like spatial navigation \citep{klautke2023dynamic, vericel2024organizing}, attention \citep{yin2012anatomical, bisley2010attention}, working memory \citep{berryhill2011intersection, hahn2018posterior, xu2024human}, and learning \citep{summerfield2020structure, whybird2021role}. We extend these insights by showing that the anterior parietal region aligns with auditory input configurations, while the posterior region aligns with visual input configurations.

Finally, while our biologically-constrained RNN successfully isolated latent configuration factors, this study has limitations, including the spatial resolution constraints of source-localized EEG, the simplified nature of the RNN model, and the use of restricted artificial stimuli. Future work should explore these patterns using naturalistic inputs and higher-resolution neuroimaging \citep{liu2025flexible, mijalkov2025computational}.

\bibliographystyle{plainnat}
\bibliography{references}

\appendix
\section{Detailed Methods}
\subsection{Participants, EEG Recording, and Preprocessing}
A total of $114$ participants ($60$ males and $54$ females, mean age $\pm$ SE = $25.2 \pm 2.4$) were recruited for the experiment to record resting-state EEG during wakefulness; among these, $77$ participants were also recorded during postprandial somnolence, and $6$ participants were recorded after $24$~h of sleep deprivation.
All participants were right-handed, low-caffeine consumers, with normal or corrected-to-normal vision and without any neurological disorders.
They were asked to get a good night's sleep before the day of the experiment and to abstain from any psychotropic substances for one day before the experiment as well as during the entire experimental procedure.
The experiment was approved by the ethics committee of the Institute of Automation, Chinese Academy of Sciences. All participants provided informed consent.

EEG data were recorded using $64$~Ag/AgCl sintered ring electrodes (EasyCap) with BrainAmp DC amplifiers (Brain Products GmbH, Gilching, Germany).
The position of electrodes was arranged according to the standard $10-20$ system. AFz served as the ground, and the reference was placed on the tip of the nose.
A vertical electrooculogram (EOG) was recorded to monitor eye movements and blinks.
The EEG signals were amplified and digitized at a sampling rate of $5000$~Hz ($0.016-100$~Hz band-pass filtering), with impedances kept below $20$~k$\Omega$.
Data acquisition was controlled through Brain Vision Recorder (version $1.03$, Brain Products GmbH, Gilching,
Germany).
For each subject and brain state, $4$-minute eyes-closed and $4$-minute eyes-open resting-state EEG were recorded.

EEG preprocessing was performed in EEGLAB~\citep{delorme2004eeglab} (version $14.1.1$) installed in Matlab (MathWorks,
Natick, MA). Raw EEG data were mapped to standard electrode locations and downsampled to $250$~Hz. The data were $0.05-45$~Hz band-pass filtered offline followed by a notch filter at $50$~Hz using a third-order, zero-phase, non-causal Butterworth filter. Artifacts from eye movements and blinks were removed using Independent Component Analysis (ICA).

\subsection{Source Localization}
We performed source localization to estimate underlying brain source signals from scalp-level EEG recordings using the Brainstorm toolbox~\citep{tadel2011brainstorm}. A three-layer symmetric boundary element model (BEM) of the head was computed using OpenMEEG, and rotating dipoles with $3003$ vertices on the cortical surface were selected. The Minimum Norm Estimation (MNE) approach, incorporating depth weighting and regularization, was applied to estimate the imaging kernel, mapping sensor-level EEG to source-level current source density. The three-dimensional current density time series for each vertex was reduced to a single dimension using Principal Component Analysis (PCA). Source-localized signals for Regions of Interest (ROIs) were extracted based on the Desikan-Killiany atlas ($68$ regions). ROI names and corresponding brain sub-networks are listed in Table S1.

\subsection{Triple Configuration Framework}
\begin{figure}
	\centering
	\includegraphics[width=\linewidth]{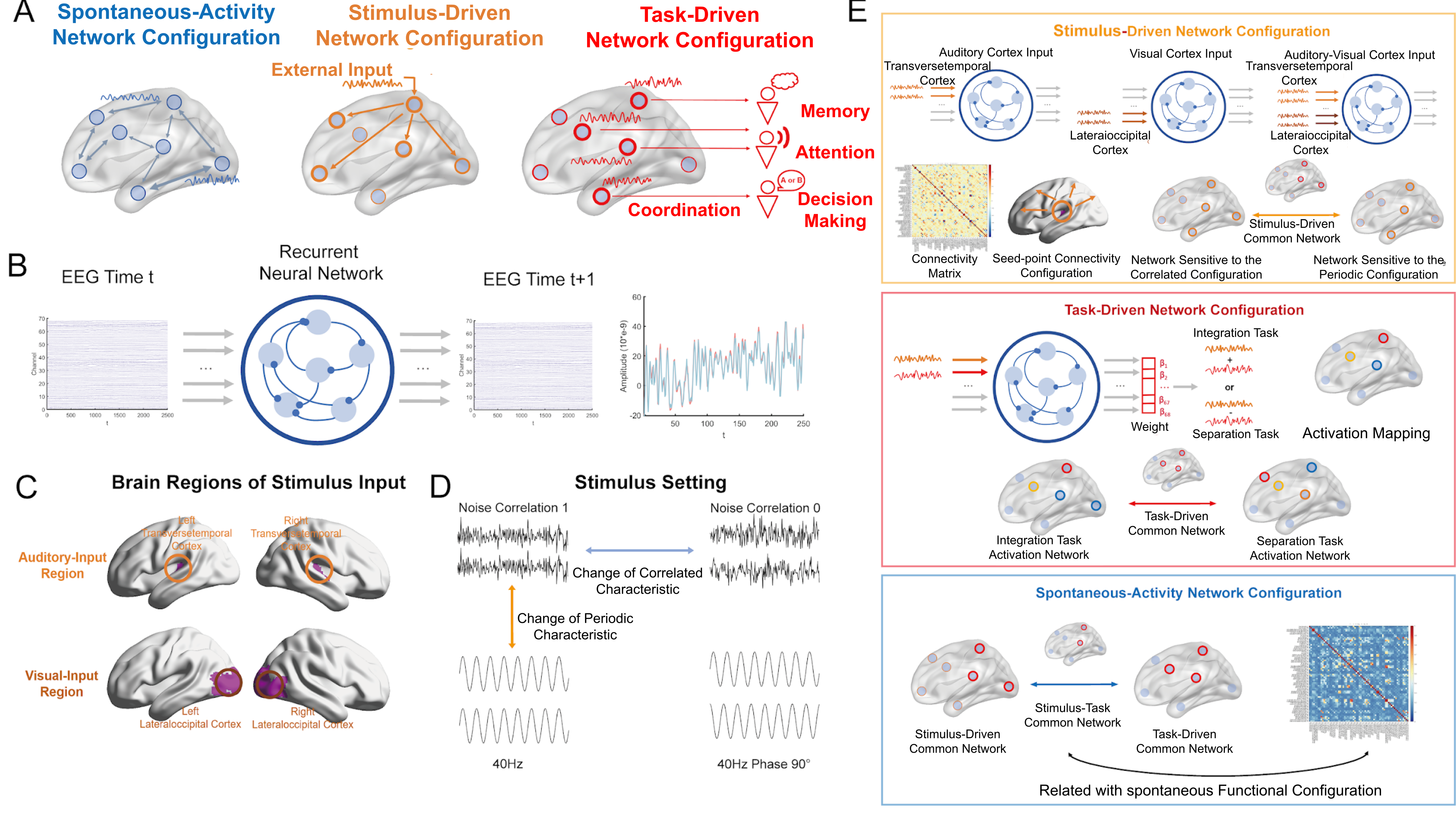}
	\caption{Study Pipeline. (A) Triple configuration framework of brain networks.
		Left: External stimulus-driven network configuration;
		Middle: External task-driven network configuration;
		Right: Internal spontaneous-activity network configuration.
		(B) RNN model training with input of source-localized EEG at time $t$ to predict EEG at time $t+1$.
		Right panel shows the predicted EEG signal (blue line) and ground truth (red line)
		(C) Brain regions of interest for sensory input. The bilateral transverse temporal regions served as auditory inputs, and the bilateral lateral occipital regions served as visual inputs.
		(D) Stimulus setting.
		(E) Analysis procedure of stimulus-driven network configuration.
		(F) Analysis procedure of task-driven network configuration. (G) Analysis procedure of spontaneous-activity network configuration.
	}
	\label{fig:overview}
\end{figure}

Specifically, the first type of configuration arises from the direct brain network response to external sensory inputs (e.g., auditory or visual) (Figure~\ref{fig:overview}A, Left); the second originates from internal information reorganization driven by task requirements (Figure~\ref{fig:overview}A, Middle); and the third stems from spontaneous brain activity, which is independent of external stimuli or tasks but intrinsically associated with the brain's state (Figure~\ref{fig:overview}A, Right). Notably, the parietal network participates simultaneously in configurations driven by external stimuli, tasks, and spontaneous activity. By comparing configuration patterns across different sensory modalities, this study further demonstrates the functional specialization of the anterior and posterior parietal networks in response to auditory and visual inputs, respectively. Collectively, these results elucidate the combined influence of endogenous and exogenous factors on brain network configurations, highlighting the parietal network's role as a central hub.

\begin{figure}
	\centering
	\includegraphics[width=\linewidth]{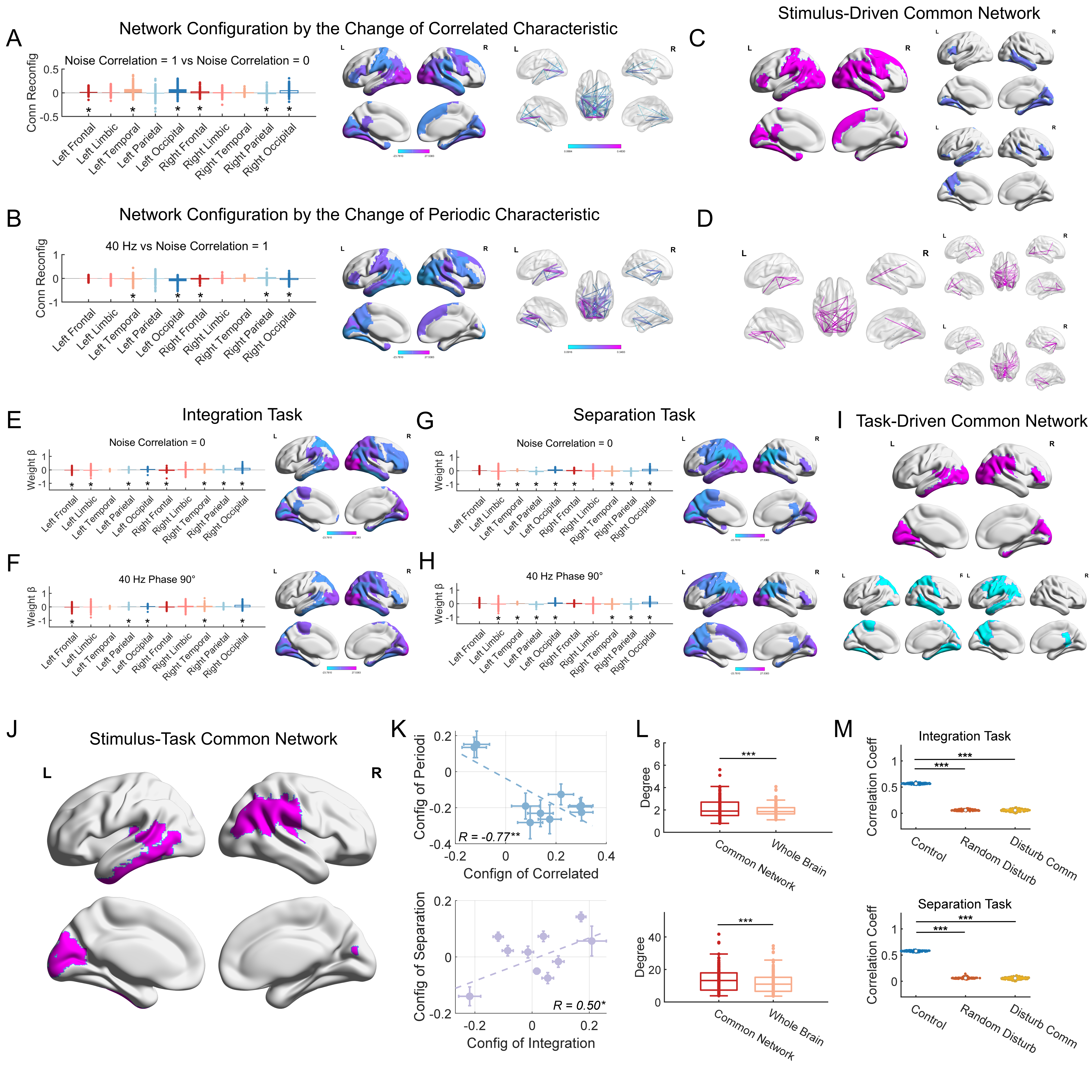}
	\caption{Triple network configuration under auditory-visual cortex input. (A)-(D) Stimulus-driven configuration. (A) Network configuration by the change of correlated characteristic (Noise correlation 1 vs Noise correlation 0). Left: seed-point connectivity configuration of sub-networks; Middle: seed-point connectivity configuration mapping on the whole brain; Right: pairwise connectivity configuration of the whole brain. (B) Same as (A) by the change of periodic characteristic (Noise correlation 1 vs 40 Hz). (C) Left: Stimulus-driven common network by combining (A) and (B); Upper-right: Network only configured by the change of correlated characteristic; Lower-right: Network only configured by the change of periodic characteristic. (D) Same as (C) shown at the pairwise connectivity level. (E)-(I) Task-driven configuration. (E) Network configuration performing integration task under the input of white noise with correlation 0. Left: seed-point connectivity configuration of sub-networks; Right: seed-point connectivity configuration mapping on the whole brain. (F) Same as (E) under the input of 40Hz with phase lag 90°. (G) Network configuration performing separation task under the input of white noise with correlation 0. Left: seed-point connectivity configuration of sub-networks; Right: seed-point connectivity configuration mapping on the whole brain. (H) Same as (G) under the input of 40Hz with phase lag 90°. (I) Upper: Task-driven common network by combining (E) - (H); Lower-left: Network only configured by the integration task; Lower-right: Network only configured by the separation task. (J)-(M) Spontaneous-activity configuration. (J) Stimulus-task common network combining (C) and (I). (K) Correlation of common network configuration patterns. Upper: different types of stimulus input; Lower: different types of tasks. (L) Comparison between the degree of brain regions in the common network and the whole-brain averaged degree. Upper: alpha band; Lower: beta band. (M) Elimination experiment. Upper: integration task; Lower: separation task. *p <0.05, **p<0.01, ***p<0.001 after Bonferroni correction. }
	\label{fig:av_triple}
\end{figure}

\begin{figure}
	\centering
	\includegraphics[width=\linewidth]{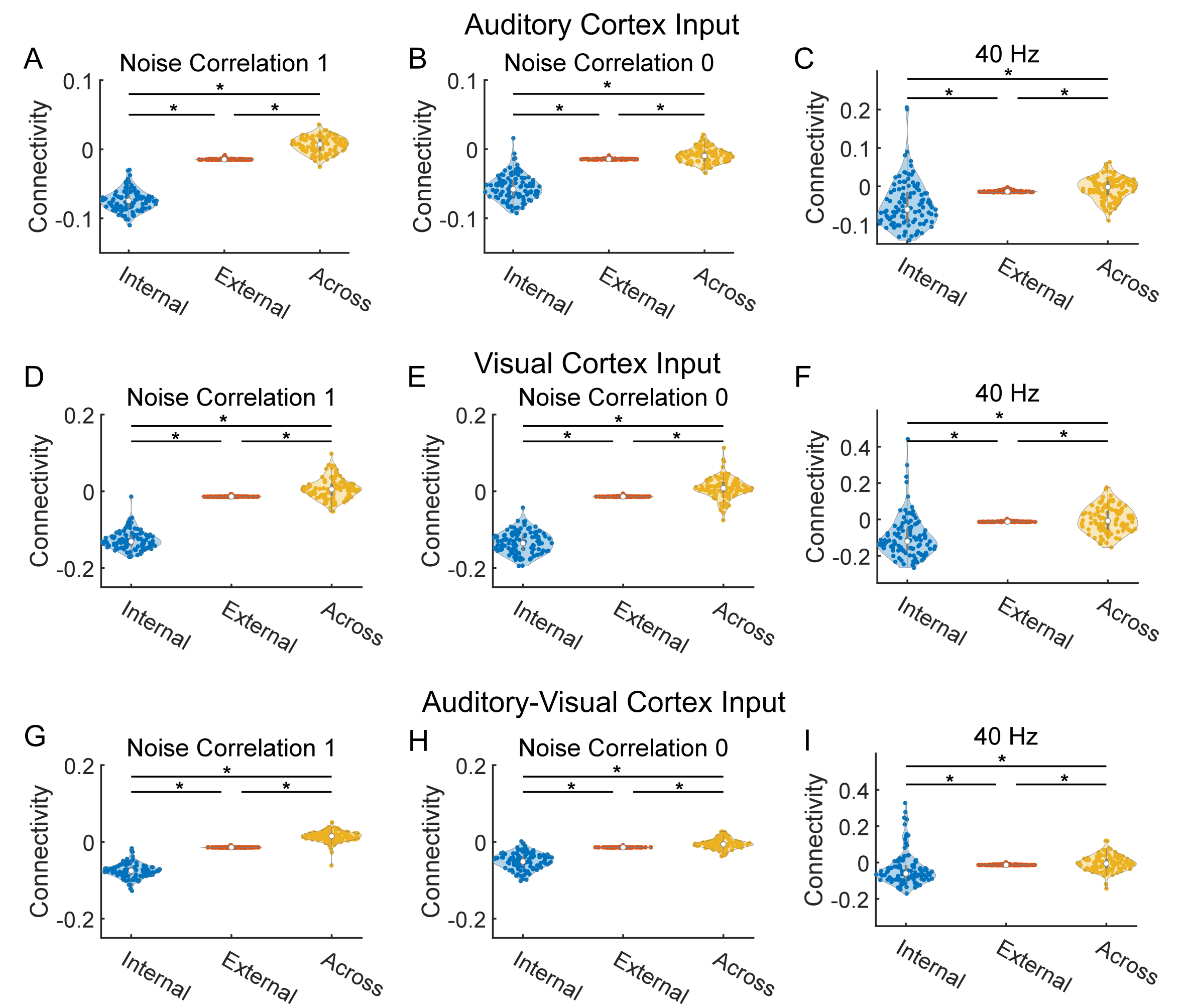}
	\caption{Comparisons of stimulus-driven connectivity among three types: connectivity within the common network (Internal ), connectivity outside the common network (External) as well as  connectivity across the internal regions and external regions of the common brain network (Across). (A)-(C) auditory cortex input; (D)-(F) visual cortex input; (G)-(I) auditory-visual cortex input. Left: input of noise correlation 1; Middle: input of noise correlation 0; Right: 40Hz. *p <0.05 after Bonferroni correction.}
	\label{fig:connectivity_comparison}
\end{figure}

\subsection{The triple-configured common network under multi-modal stimuli recruited the common network of the single-modal stimuli}

We then investigated network configuration under simultaneous auditory and visual input. Stimuli were redefined: aperiodic white noise perfectly correlated ($r=1$) or uncorrelated ($r=0$) between auditory and visual cortices; and periodic $40$~Hz signals with $0^\circ$ or $90^\circ$ phase lag between modalities. Stimulus-driven configuration (Figure~\ref{fig:av_triple}(A)-(D)) revealed that correlation and periodicity changes, alongside the resultant common network, were distributed across the frontal, temporal, parietal, and occipital networks (Figure~\ref{fig:av_triple}(A)-(C), t-test, $p<0.05$, Bonferroni correction). Pairwise analysis confirmed predominant involvement of intra-temporal, parietal, and occipital connections (Figure~\ref{fig:av_triple}(A)-(B) Right, (D) Left, t-test, $p<0.05$, Bonferroni correction).

Task-driven configuration demonstrated widespread regional participation in both integration and separation tasks (Figure~\ref{fig:av_triple}(E)-(H), t-test, $p<0.05$, Bonferroni correction). The task-driven common network localized to temporal, parietal, and occipital regions (Figure~\ref{fig:av_triple}(I), Upper). The combined auditory-visual stimulus-task common network (Figure~\ref{fig:av_triple}(J)) integrated regions from both the auditory common network (supramarginal, bankssts, inferiortemporal) and visual common network (inferiorparietal, cuneus, pericalcarine). Dynamically, correlation configuration was negatively correlated with periodicity configuration (Pearson $R = -0.77, p<0.01$, Figure~\ref{fig:av_triple}(K), Upper), while integration and separation activations were positively correlated (Pearson $R = 0.59, p=0.04$, Figure~\ref{fig:av_triple}(K), Lower). Consistent with single-modal results, this multi-modal common network exhibited significantly higher degree in alpha and beta PLV networks (Figure~\ref{fig:av_triple}(L)), indicating spontaneous activity coupling. Eliminating this common network significantly impaired performance in both integration (Figure~\ref{fig:av_triple}(M), Upper, One-way ANOVA, $p<0.001$; post-hoc paired t-test, $p<0.001$, Bonferroni correction) and separation tasks (Figure~\ref{fig:av_triple}(M), Lower, One-way ANOVA, $p<0.001$; post-hoc paired t-test, $p<0.001$, Bonferroni correction). Thus, multi-modal network configuration recruits single-modal hub networks.

\subsection{Local connectivity within the common network was strengthened under the corresponding external input}
We further evaluated connectivity within the common network (Internal), outside the common network (External), and across these boundaries (Across) under auditory (Figure~\ref{fig:connectivity_comparison}(A)-(C)), visual (Figure~\ref{fig:connectivity_comparison}(D)-(F)), and multi-modal (Figure~\ref{fig:connectivity_comparison}(G)-(I)) input (One-way ANOVA, $p<0.001$; post-hoc paired t-test, $p<0.05$, Bonferroni correction). Across all input types, internal connectivity patterns were consistent, characterized by significant anti-correlated connections. Spontaneous functional connectivity (alpha/beta PLV) was significantly higher inside the common networks compared to outside (Figure S4(A)-(F), One-way ANOVA, $p<0.001$; post-hoc paired t-test, $p<0.05$, Bonferroni correction), implying inherent differentiation to support triple configuration. Specifically, alpha/beta PLV inside the auditory common network exceeded that of the visual and multi-modal common networks (Figure S4(G)-(H), One-way ANOVA, $p<0.001$; post-hoc paired t-test, $p<0.05$, Bonferroni correction).

\end{document}